\documentclass[10pt]{iopart}

\usepackage{amssymb}
\usepackage{graphicx}
\usepackage{epstopdf}
\usepackage{textcomp}
\usepackage{color}
\usepackage{filecontents}
\usepackage{cite}
\usepackage{url}
\usepackage{iopams} 
\expandafter\let\csname equation*\endcsname\relax
\expandafter\let\csname endequation*\endcsname\relax 
\usepackage{amsmath}
\usepackage{amsfonts} 

\begin{document}

\title[Confined H$_2^+$]{Monte Carlo Calculation of the Potential Energy Surface for Octahedral Confined H$_2^+$}
\author{Savino Longo$^{1,2}$, Gaia Micca Longo$^{1}$ , Domenico Giordano$^{3}$ }

\address{$^1$ Department of Chemistry, University of Bari, via Orabona 4, 70126 Bari, Italy}
\address{$^2$ CNR-Nanotec, via Amendola 122/D, 70126 Bari, Italy}
\address{$^3$ ESA-ESTEC, Aerothermodynamics Section, Kepleerlaan 1, 2200 ag, Noordwijk,
              The Netherlands}

\ead{savino.longo@uniba.it, gaia.miccalongo@uniba.it, domenico.giordano@esa.int}

\vspace{10pt}
\begin{indented}
\item[]October 2017
\end{indented}

\begin{abstract}

A rich literature has been produced on the quantum states of atoms and molecules confined into infinite potential wells with a specified symmetry. Apart from their interest as basic quantum systems, confined atoms and molecules are useful models for extreme high pressure states of matter, spectroscopically active defects in solid lattices and chemical species in molecular cages. A most important case is that of H$_2^+$ for which little or no results are available in the case of polyhedral confinement. The approach of the authors makes use of the Diffusion Monte Carlo (DMC) method. The advantage of this method is that previously developed codes are readily adapted to new, even complex, well geometries and nuclear positions. In this paper the potential energy surface (PES) of H$_2^+$ confined inside an octahedral well is reported for restricted D$_{4h}$ and D$_{3d}$ geometries and different well widths. The results are discussed using the concept of electron compression and the correlation with semi-confined atomic orbitals.
\end{abstract}

\ioptwocol

\section{Introduction}
\label{intro}

A rich literature has been produced on the quantum state of atoms and molecules confined into potential wells with a specified symmetry not only as basic quantum systems but also as models for extreme high pressure states of matter, spectroscopically active defects in solid lattices and chemical species in molecular cages (an updated review of the state of the art is found e.g. in \cite{leykoo2014} while applications are discussed e.g. in \cite{Raman}).
As expected much attention has been devoted to H$_2^+$, the simplest molecular system\cite{cruz2009,colin2011,leykoo1981,sarsa2012,mateos2002,gorecki1988,singh1964,micca20151,micca20152,micca20153,segal2006}.
Also this particular case, apart from its value as fundamental issue, may find several applications.
For example, it was studied as a model system for a molecular qubit in a solid \cite{kang2008} and as a reaction intermediate of electrochemical reactions involving molecular hydrogen on Pt electrodes \cite{juodkazis2011,juodkazyte2014}.
Confined H$_2^+$ also represents the simplest model for a F$_2^+$ center in crystals \cite{henderson2006} and these ions are actually implanted into solids (e.g. KBr \cite{KBr}) purposely.
Confined H$_2^+$ presents also an interest for materials science and microelectronics \cite{micca20153}, being the probable outcome of H$_2^+$ ions impact on solid substrates, which are charged at high negative voltage with respect to the plasma potential in a radio-frequency or microwave reactor with H$_2$ as main component of the feed: these devices have attracted much attention in the last few years also for their use in materials processing \cite{hydro1,hydro2,hydro3, hydro4,diomede2012,diomede2014,diomede2017}. Although previous studies have shown that H$_3^+$ is the main ion present in the plasma phase under typical pressure conditions for these devices, the triatomic ion dissociates above 4.37 eV and it is not found inside solid lattices after exposure to plasma \cite{otte1998}. These results are also potentially relevant for studies of fusion plasma devices and future fusion reactors walls, constantly incorporating high energy D$^+$, T$^+$ ions \cite{iter1,iter2}.

Since very few information is found on the physical state of H$_2^+$ inside real solid lattices, simple geometric models of confinement are very appealing as a framework, or physical environment, to standardize preliminary calculations and compare results.

The confined species was found to possess different properties with respect to the free one. It has been shown that confined H$_2^+$ in its most stable electronic state ($^2\Sigma_g^+$) has a lower bond length due to electron cloud compression and that the related Potential Energy Surface (PES) has a different shape. The compressed states corresponding to the unbound $^2\Sigma_u^+$ and $^2\Pi_g$ present deep minima in their potential curves \cite{micca20151,micca20153}.

Some results have been reported for non Born-Oppenheimer calculations of confined hydrogen systems \cite{skouteris2010,sarsa2012}. However, the availability of PESs for the classical motion of nuclei of H$_2$/H$_2^+$ into potential wells discloses the wide field of vibrational level calculations and prediction of Raman spectra to be compared to experiments and {\it ab initio} calculations. 

In spite of the very wide perspective for further studies and the very high rate of results production in the last years, as mentioned, previous works have, as a rule, considered only spherical and spherical prolate confinement. Cylindrical confinement has been considered for H$_2$ \cite{lo2005}, cylindrical and cubic confinement for the H atom \cite{micca20152}.

In this work, we report results for H$_2^+$ confined in a octahedral well using a slightly modified version of the Monte Carlo code used by the same authors in the past.

\section{Calculation method}\label{method}
The method used for these calculations has been already discussed in previous works by the same authors \cite{micca20151,micca20152,micca20153}. It can be described as a diffusion Monte Carlo (DMC), with some differences with respect to the standard one, in order to achieve adaptability to different confining polyhedra. Basically, DMC uses a diffusion process in a phase space to project an initial guess for the wave function onto the lowest energy state with a given symmetry. During the propagation along imaginary time, the energy guess is adapted. Although the DMC method is usually not considered the best choice for a few particles problem, we have shown in our previous papers that is leads to straightforward formulation of numerical codes for any shape of the well. Furthermore, Cartesian coordinates can be used for any new case without necessity to match the geometry of the well \cite{micca20152}. Algoritms produced this way have been validated by comparison with known results \cite{micca20152}). Concerning numerical and algorithm implementation issues, they have been described in much detail elsewhere \cite{micca20153}. Accordingly, only the additional implementation issues specific to the application of the method to an octahedral well are explained in the next section.

\section{The octahedral well}\label{OCt}

\begin{figure}
\centering
\resizebox{0.5\textwidth}{!}{
  \includegraphics{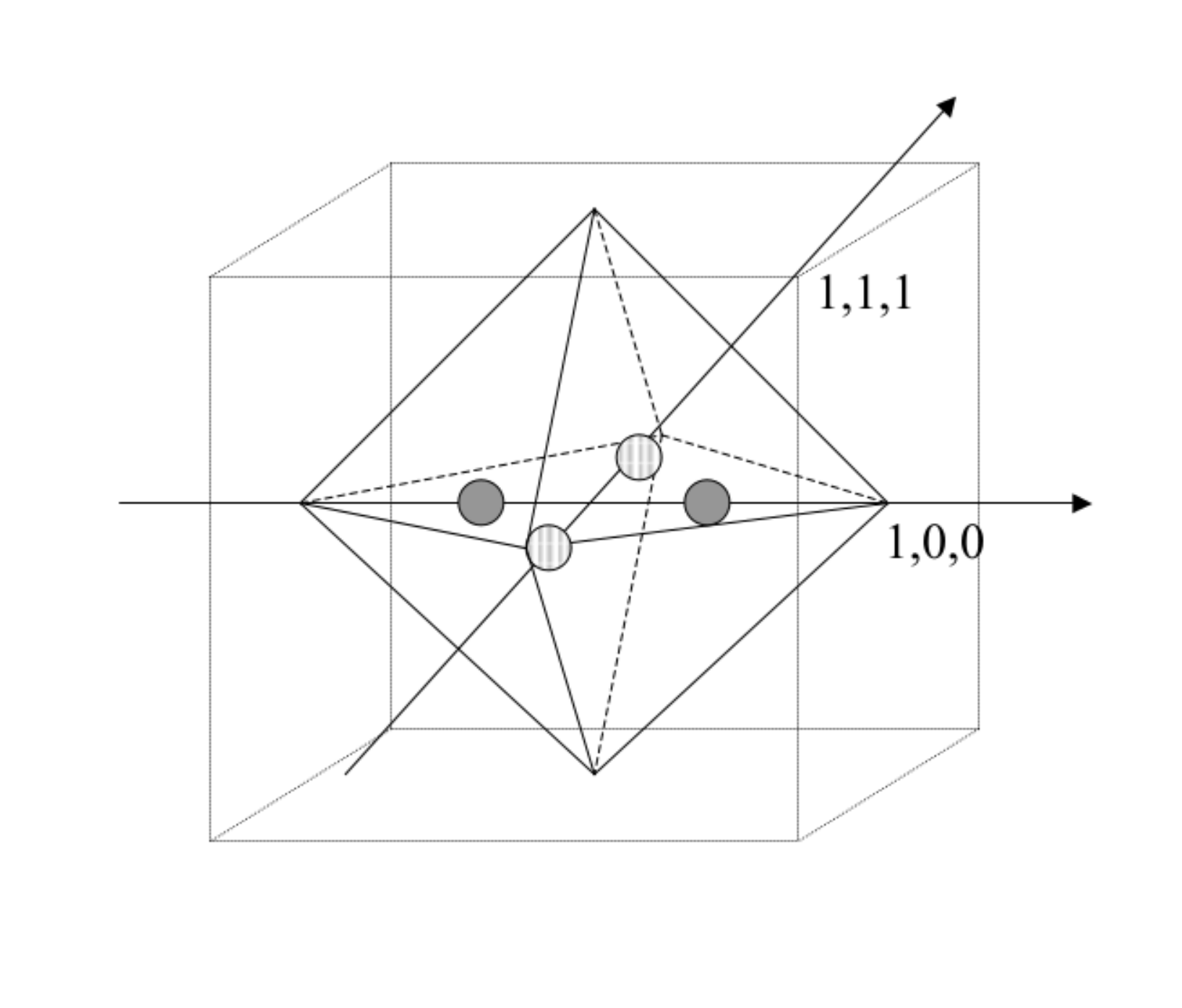}
}
\caption{Octahedral confinement of H$_2^+$. The plot displays nuclear and barrier locations, used in the calculation of the electronic eigenfunction, in the two cases considered in the paper. In the plot the confining walls are related to a cubic lattice in order to serve as an analogy of a octahedral hole. Related Miller indexes used in the calculations are shown. The two cases, as shown, are [100], where the two nuclei are placed on a C$_4$ axis of the O$_h$ well giving a D$_{4h}$ symmetry, and [111] where they are placed on a C$_3$ axis giving a D$_{3d}$ symmetry.}
\label{fig:1} 
\end{figure}

The most relevant symmetry coordinates for a diatom inside an octahedral well are [100] (using Miller's notation)  corresponding to a diatom on a C$_4$ axis and [111]  corresponding to a diatom on a C$_3$ axis (Figure \ref{fig:1}).

The two cases possess a reduced symmetry with respect to initial D$_{\infty h}$ of the isolated molecule and O$_{h}$ of the empty potential well, being respectively D$_{4h}$ for [001] case and D$_{3d}$ for [111] . 

These directions are along $O_h$ symmetry axes of the barrier.

We assume the following Hamiltonian:

\begin{equation}\label{1}
\hat{H}=-\frac{1}{2}\nabla^2+V(\textbf{r})+\frac{1}{d}+E_H
\end{equation}

that includes the repulsion energy of the two nuclei, which is classical in nature. The addition of $E_H$, which is the ionization energy of the $1s$ state of H, fixes $E_T=0$ as the asymptotic value of a free (confinement dimension $r_0 \rightarrow \infty $) ion at large $d$ (dissociation limit) in the  ground electronic state.

$V({\bf r})$ is the expression of the potential energy due to nuclei and barrier on the electron, i.e. 

\begin{equation}\label{2}
V({\bf r}) = -\frac{1}{|{\bf r}-{\bf r}_a|}-\frac{1}{|{\bf r}-{\bf r}_b|}+V_B({\bf r})
\end{equation}

In this expression ${\bf r}_a$ and ${\bf r}_b$ are the position vectors of the two nuclei, which are expressed as a function of $d$ after the Miller index of the internuclear axis $\left [ ijk \right ]$ are fixed. $V_B$ is the barrier contribution, which equals zero inside the polyhedral well and becomes indefinitely high outside.

With the present choice of the axis, the position of first nucleus is given as a function of $\left [ ijk \right ]$ and $d$ by

\begin{align}\label{3}
x_a = d \frac{i}{2\sqrt{(i^2+j^2+k^2)}}\\
y_a = d \frac{j}{2\sqrt{(i^2+j^2+k^2)}}\\
z_a = d \frac{k}{2\sqrt{(i^2+j^2+k^2)}}
\end{align}

The second nucleus is then at ${\bf r}_b=-{\bf r}_a$.

Octahedral confinement is obtained by using the following expression of $V_B$:

\begin{equation}\label{4}
V_B({\bf r}) = 
   \begin{cases}
    0       & \quad \text{if } |x|+|y|+|z|<c\\
    \infty  & \quad \text{if } |x|+|y|+|z|>c\\
  \end{cases}
\end{equation}

where $c$ is the distance between the center and any vertex of the octahedral hole.

In a cubic lattice $2c \sim a$ where $a$ is the lattice constant. However, the exact relation between $c$ and $a$ can be established only on the basis of a fitting, which requires a spectroscopic quantity to be determined and compared to experiments for H$_2^+$ in a given material, e.g. a Raman spectrum. 

\section{Results}\label{Results}

For both cases, we give here only the PES for lowest a$_{1g}$ orbital, although excited orbitals can be easily calculated as shown in \cite{micca20151,micca20152,micca20153}

\begin{figure}
\centering
\resizebox{0.5\textwidth}{!}{\includegraphics{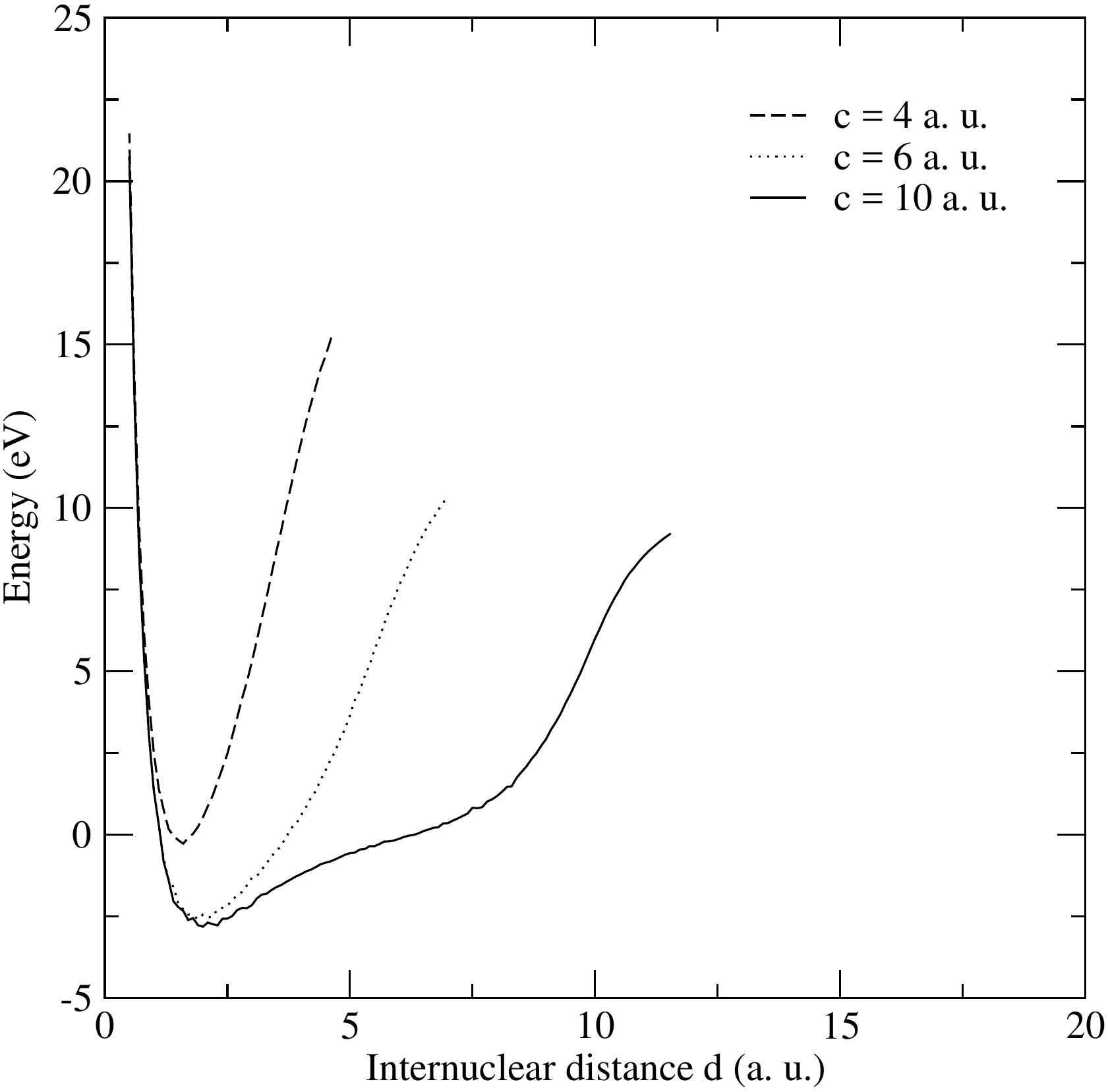}}
\caption{Octahedral confinement: [111] direction (D$_{3d}$). H$_2^+$ potential energy as a function of internuclear distance $d$.}
\label{fig:2} 
\end{figure}

\begin{figure}
\centering
\resizebox{0.5\textwidth}{!}{\includegraphics{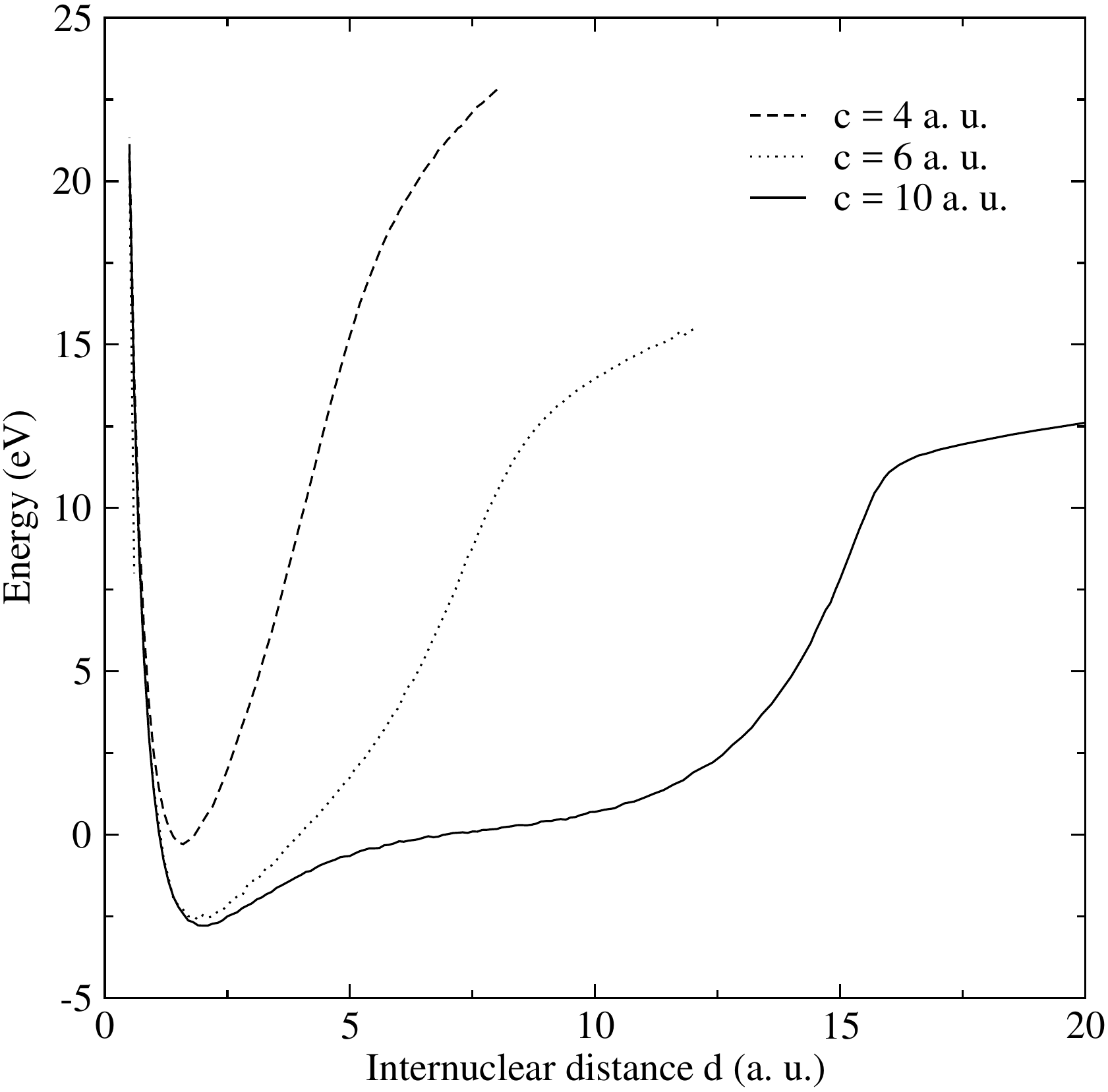}}
\caption{Octahedral confinement: [100] direction (D$_{4h}$). H$_2^+$ potential energy as a function of internuclear distance $d$.}
\label{fig:3} 
\end{figure}

Results are given here for three values of the parameter $c$, namely 4, 6 and $10 a.u.$ The first two are of the right order of magnitude to be relevant for applications while the largest one is considered in order to discuss the limiting behavior for large wells. Results are reported in tables \ref{Tab1} and \ref{Tab2} and in figures \ref{fig:2} and \ref{fig:3}. These results show that the physical state of the ion in a well of fair dimension to mimic an octahedral hole is very different from that in plasma phase. For example, the compression in a well with $c = 4 a.u.$ changes drastically the energy of the minimum of the PES, its curvature close to the minimum and the equilibrium distance between the two nuclei (Table \ref{Tab1}). Also, in agreement with previous results for confined H$_2^+$ also by the present authors \cite{micca20151,micca20153}, it is observed that the confinement produces a strong increase of the energy for values of $d$ larger than the equilibrium one, leading to a peculiar shape of the PES corresponding to a strongly bound state with high vibrational frequency. 

An issue that is due to investigate here, since it was not relevant in previous calculation, is the effect of the orientation of the internuclear axis relative to the barrier. It has been found in the past that the confinement of a molecule into a well with dimensions comparable to the molecular radius increases the vibrational frequencies and leads to "bound-like" PESs also for originally non-bound states. The effect is due to a mechanism which can be interpreted as due to "compression" of the electron "cloud" increasing the negative potential terms when the nuclei are displaced \cite{micca20151}.

A different mechanism, which occurs in cases where $d$ is much larger than the molecular radius, discussed in \cite{micca20153}, is of much interest here. It can be seen that for the case $c = 10 a.u.$ the molecule has enough room to practically dissociate, and the PES gets close to the zero value at intermediate $d$ values; al larger $d$ however, the energy increases and reaches a maximum when the nuclei are in contact with the barrier (full lines in Figs. 2 and 3).

The basic of this mechanism is that the barrier surface is a nodal one. Its effect can be discussed in terms of symmetry. For example, in a hydrogen atom close to a dielectric surface, the $1s$ atomic state is symmetry correlated to the $2p$ state, which has the same number of radial nodes, when the nucleus is in contact with the surface. This leads to a theoretical energy of $10.2$ eV. Indeed a close energy is obtained, as expected, for the D$_{3d}$ case of  figure \ref{fig:2} when the well is large enough (see also Table \ref{Tab2}).

Instead, in figure \ref{fig:3} it is possible to note that the PES varies with $d$ by a much different trend when nuclei are approaching the vertex and it reaches a limit energy not far from the H ionization one (Table \ref{Tab2}). This region corresponds to a limit hydrogen atom inside the octahedral edge where four octahedral faces join, so the electron cloud is strongly compressed. The limit atom of the H$_2^+$ dissociation is therefore in a highly excited state, close to ionization.

\begin{table}

\begin{tabular}{|c|c|c|c|}
 \hline
  
  & c (a.u.) & d$_{eq}$ (a.u.) & E$_{eq}$ (eV)\\
 \hline
 
 
 
 [100] & 6 & 1.9 & -2.60\\
 
 [111] & 6 & 1.9 & -2.59\\
 \hline
 
 [100] & 4 & 1.6 & -0.29\\
 
 [111] & 4 & 1.6 & -0.27\\
 \hline

\end{tabular}
\caption{Energy values at equilibrium distances, for [100] and [111] directions.}
\label{Tab1} 
\end{table}

\begin{table}

\begin{tabular}{|c|c|c|c|}
 \hline
  
  & c (a.u.) & d (a.u.) & E (eV)\\
 \hline
 
 [100] & 10 & 20 & 12.61\\
 
 [111] & 10 & 11.55 & 9.22\\
 \hline
 
 [100] & 6 & 12 & 15.46\\
 
 [111] & 6 & 6.93 & 10.22\\
 \hline
 
 [100] & 4 & 8 & 22.81\\
 
 [111] & 4 & 4.62 & 15.21\\
 \hline

\end{tabular}
\caption{Energy differences between [100] and [111] directions.}
\label{Tab2} 
\end{table}


\section{Conclusions}\label{conc}
In this paper, the confinement of a H$_2^+$ ion in an octahedral well is considered for the first time. From the point of view of plasma devices and materials science, this system represents a model for H$_2^+$ into a octahedral hole of a solid lattice, where the real hole is replaced  by an infinite  well with the same symmetry and behaving  like a three dimensional  box  for  the  electron.  Orbitals  are  determined by  solving  the  Schroedinger  equation  using a version of the DMC method already applied by the authors to  confined H  in various geometries and spherical confined H$_2^+$. The PESs as a function of $d$ for two different orientations, [001] and [111] are reported for a few values of the well width representative of  solids, in particular Pt for which confined H$_2^+$ has been considered in the past as reaction intermediate. Results confirm the known phenomenology of confined H$_2^+$, but allow a further discussion in view of the role played by molecular orientation relative to the well shape. In perspective, ideal octahedral confinement obtained by rigid walls, analogous to the very much studied spherical, ellipsoidal and cylindrical, may provide a very useful reference model for initial guesses of spectroscopic predictions to be compared with e.g. Raman or optical absorption measurements, thereby providing useful information on the physical status of molecular species retained into solid lattices with account of the real lattice symmetry, in particular species produced from the penetration of high energy particles from a hydrogen plasma into reactor wall materials and substrates. In this first paper on the topic, we have not addressed the issue of excited orbitals with known nodal surfaces which are promptly accessible using our method. Future studies could also consider less restricted geometries where the molecular center of mass is not in the center of the well. Such extension presents, even prior to actual calculations, interesting symmetry aspects to start future works with.

\section*{Acknowledgment}
This research activity has been supported by the General Studies Programme of the
European Space Agency under 
grant 4200021790CCN3. S.L. acknowledges useful e-mail discussions with C. Le Sech.

\end{document}